# Did bio-homochirality arise from spin-polarized electron?

Wei Wang*

School of Physics, Harbin Institute of Technology, Harbin, China

Email: wwang_ol@hit.edu.cn

**Abstract**: The origin of bio-homochirality is a subject of much debate. The emergence of chirality and life on earth is a break of symmetry to be compared with the breaks of symmetry in the evolution of the universe. Based on a perspective of asymmetry transfer, the chirality at molecular level might stem from electron spin at subatomic level. Accordingly, in this paper a spin-induced chiral selectivity (SICS) mechanism and its outreach are introduced and discussed. The stress force or spin torque derived from quantum electrodynamics (QED) might be the driving force for the transfer of asymmetry and the formation of molecular chirality. Some recent experimental results seem to support the SICS conjecture. If spin-polarized electrons (SPEs) did cause life to become chirally selective, a magnetic half-metal material such as greigite ($Fe_3S_4$), a mineral present in a primordial site where life could have emerged, might act as a spin filter to produce SPEs, which then induced the asymmetric synthesis of chiral molecules via the SICS mechanism. All these tentative thoughts may help explain how homochirality and life could have arisen on the early Earth.

## Introduction

Harold Morowitz at George Mason University preached a simple philosophy: if you want to understand the chemical emergence of life, look to life's most basic chemistry—the biochemical pathways shared by all organisms [1]. Chirality is such a commonly-shared feature of all living organisms on Earth. A chiral molecule with one chiral center can be resolved into two enantiomers with nonsuperimposable mirror images. However, life on Earth consists almost exclusively of one enantiomer, for example, L-amino acids and D-sugars. This is the so-called bio-homochirality, one of the key challenges in the origin-of-life science.

Biologically active polymers such as proteins and nucleic acids are composed of homochiral building blocks. As a result, the α-helixes of protein and the double strands of DNA in their standard forms always twist like a right-handed screw. Only in this way can life execute its biochemical functions and life cycle exactly. That is to say, no chirality, no structure, no function, and then no life. However, how did the bio-homochirality form in the prebiotic world still remains a giant mystery to date.

In a couple of early papers [2-4], I have suggested a tentative scenario for the question: magnetic half-metal materials such as greigite ($Fe_3S_4$), a mineral present in a possible primordial site where life could have arisen, might act as a spin filter to produce spin-polarized electron (SPE); the resulted SPEs could then induce the asymmetric synthesis of chiral organic molecules via a quantum mechanical effect. This is called *spin-induced chiral selectivity* (SICS).

In the previous papers [2-4], the SICS conjecture was palletized and loosely introduced. As of today, some prima facie evidence supporting the conjecture has made its first appearance. Therefore, in this paper, after a brief revisit to the SICS mechanism, a wider array of outreach considerations will be presented, as well as a review on those recent experimental achievements. All these may help uncover the possible role of SPE in the origin of bio-homochirality.

## SICS mechanism and outreach

In the two earlier papers [2, 4], I have proposed a reaction model to validate the SICS mechanism (Fig.1),

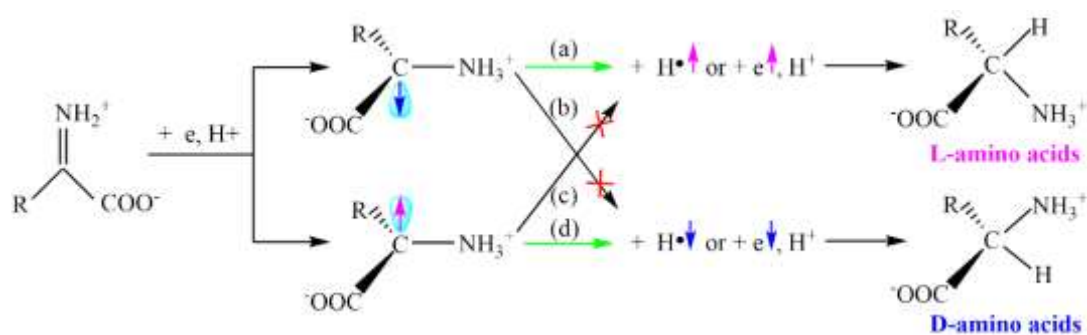

**Fig.1** Schematic illustration of the SPE induced stereoselective synthesis of α-amino acids.

As illustrated in the above stepwise reaction, the reduction of the imine molecule needs the transfer of two electrons, always proceeding in two successive univalent steps. Firstly, an electron and a proton are added to the positive iminium ion, producing a carbon-centered amino acid radical intermediate. Then in the second step, another set of proton and electron may attack the radical sequentially or concertedly, both called proton-coupled electron transfer (PCET) [5]. The concerted pathway is considered kinetically advantageous over the sequential pathway and referred to as concerted electron-proton transfer (CEPT) or concerted PCET. The SICS reaction in Fig. 1 occurs by CEPT because both the electron and proton are transferred to the same newly-formed C-H σ bond. If the electron and proton work as an integral entity, the reaction would be termed hydrogen atom transfer (HAT), a sub-class of CEPT. In fact, the electrocatalytic reduction of a C=X unsaturated bond (where x is $CR_2$, NR, or O) always takes place in this way, since the reductive species generated on the surface of an electrode is a hydrogen atom but not an electron [6, 7].

In the second step, there are four possible spin state combination forms between the radical and the H-atom since they both have two spin states. According to the Pauli

Exclusion Principle, however, two electrons in a bonding orbit can never have the same spin. Therefore, pathways (b) and (c) in Fig. 1 are forbidden. The hypothesis is that pathways (a) and (d) might produce two mirror images of an amino acid. If channel (a) brought about L-amino acids, channel (d) would yield D-amino acids.

Chirality is not exclusive to biomolecules. In a more general sense, it could be thought of as an intrinsic property of natural objects of various scales, from elementary particles, organic molecules, biological structures, to helicoidal nebula in the universe. The mesoscopic molecular chirality at molecular level in chemistry may be determined by the microscopic electron spin at subatomic level in physics. Such a single handedness feature can be transmitted downstream and manifested in the macroscopic biological asymmetry, e.g., the shell chirality of snails, at individual level in biology (Fig. 2).

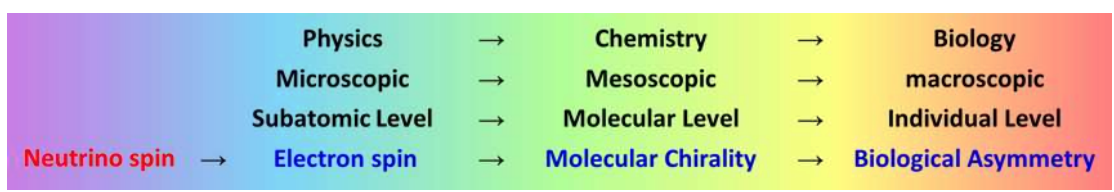

**Fig. 2** The suggested cascading chirality transfer across different scale ranges.

On the other side, the electron spin may arise from the asymmetry of neutrino spin. When an elementary particle is detected, its spin must be in one of two possible states. The spin of a right-handed particle points along the particle's direction of motion, while the spin of a left-handed particle points opposite to the particle's direction of motion. In the Standard Model of physics, a chiral twin has been found for every matter and antimatter particle, but with the exception of neutrinos. Till now, only left-handed neutrinos and right-handed antineutrinos have ever been observed. If no right-handed neutrinos exist, left-handed neutrinos are naturally chiral. A neutrino is neutral and extremely tiny, travels near the speed of light, very rarely interacts with matter, and then is very hard to be detected and researched experimentally. It is now generally considered that the absence of right-handed neutrinos could be responsible for another symmetry breaking fact that everything is made almost entirely of matter over anti-matter in the universe. Neutrinos are born and survive only with left-handed chirality.

Therefore, the symmetry breaking events might have taken place at the beginning of the universe instantaneously after the Big Bang. Neutrinos were born with a single spin state. Their chirality transmitted from themselves to electrons, molecules and then bio-entities, and manifested in different chiral forms (e.g., spin states and enantiomers) at different scale levels. The emergence of life might be a stepwise process in the fight for new highs of asymmetry and order (Fig. 2).

From the viewpoint of classic mechanics, force is an interaction between two or more objects that when unopposed causes a change in the state of motion of the objects. By a logical extension of this force-motion philosophy, the stepwise transmission (motion) of chirality across different scale ranges also needs the action of forces (Fig. 2). For example, both the double strand structure of DNA and the α-helical structure of proteins are formed mainly by hydrogen bond interaction, which in essence is an electromagnetic force, one of the four fundamental forces of nature. The other three are the gravitational force, the strong nuclear force, and the weak nuclear force. Then, among these four fundamental forces that rule the physical world, which one can induce the chiral discrimination between the two spin states of a neutrino or an electron, and then the two enantiomers of an organic molecule?

One of the most fundamental laws concerning symmetry in physics is the Noether's Theorem, which states that every physical symmetry leads to a conservation law. For example, time-translation symmetry implies the conservation of energy, and space-translation symmetry implies the conservation of momentum. Physicists once deeply believed that conservation laws were generic rules governing the world. Besides the kinematic symmetries of space-time translations, permutation symmetry underlying the statistical behavior of distinguishable fermions and bosons, and gauge symmetry describing the particle fields, physicist also often talk about three mirror symmetries in nature: charge conjugation (C), space inversion (P, parity) and time reversal (T). The CPT symmetry/conservation means that if you flip a particle to its antimatter twin (e.g., an electron into a positron), or make it right-handed instead of left-handed, or move it backward through time instead of forward, that particle should still behave in the same way and obey the same laws as it did before getting flipped.

In 1956, the Nobel Prize laureates Tsung-Dao Lee and Chen-Ning Yang suggested that parity might not be preserved under the action of the weak force [8]. The proof of parity nonconservation has since been observed in β[-] and β[+] decay experiments, by measuring the angular distribution of the emitted electrons and positrons [9, 10]. In β[-] decay of a radioactive nucleus in which the number of neutrons is excessive, a neutron (n) transforms into a proton (p) with the emission of an electron (e[-]) and an electron antineutrino ($\bar{\nu}_e$).

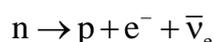

$$n \rightarrow p + e^- + \bar{\nu}_e$$

If parity is conserved, the electrons will come out of the $^{60}$Co nuclei in two opposite directions. However, Wu *et al.* found that the electrons only came out in one direction, implying the electrons were spin-polarized with only left-handed chirality, the concomitant antineutrinos were right-handed, and the parity was violated. A couple of later contributions further demonstrated that in weak interactions both C and P in CPT symmetry are imperfect, called CP violation [11-14]. But most physicists think CPT symmetry still holds at the particle level, and no particle has been found that breaks all three elements of this symmetry.

The two mirror images of a chiral biomolecule show stereosymmetry of parity. Life on Earth adhibits biomolecules of only a certain handedness. According to the forementioned force-motion philosophy, what force could have driven this selection before the origin of life on the early Earth? Among the four fundamental forces in nature, gravity, electromagnetic force and strong force are ambidextrous, treating particles equally regardless of their handedness. Only the weak force plays favorites, making the parity violated. The weak force via the Electroweak Theory renders all atomic nuclei chiral, and consequently all atoms and molecules own different degrees of chirality, regardless of whether they possess stereogenic atoms or not [15]. This cascading symmetry breaking might further influence the origin of the preferred handedness in biomolecules such as L-amino acids and D-sugars, and then the right-handed DNA double strands and protein α-helixes.

However, the weak interaction has the shortest range of action of the four fundamental forces, only participating in nuclear reactions at quark to nucleon's scales ($10^{-18} \sim 10^{-15}$ m), such as the β⁻ decay. In contrast, it is common sense that chemical reactions for molecular synthesis take place at atomic to molecular scales ($10^{-10}$ m). The range of the weak force is extremely short. How can it transmit the intranuclear parity violation information outside, and then transform into the symmetry breaking of chiral molecules, over a distance of several orders of magnitude larger than its scope? And who is the messenger?

A spin polarized electron may play such a role. Taken the β⁻ decay as an example, a proton in the nucleus turns into a neutron, emitting an electron and an antineutrino.

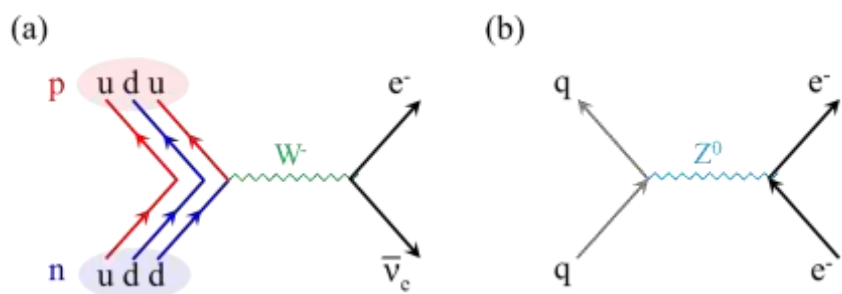

**Fig. 3** Feynman diagrams for the weak interactions involved in β⁻ decay.

During this process, the weak force changes a down quark (d) into an up quark (u). At the point of change, a gauge boson W⁻ is emitted, which promptly (within about $10^{-25}$ s) decays into an electron and an electron anti-neutrino (Fig. 3(a)). In its counterpart β⁺ decay, a 'u' quark in a proton is converted into a 'd' quark to form a neutron, releasing a gauge boson $W^+$ which very quickly degenerates into a positron and an electron neutrino. According to the Electroweak Theory, in weak interaction there is a third uncharged gauge boson $Z^0$, which mediates the exchange interaction between an electron and a quark (q) or nucleon asymmetrically, without changing their nature (Fig. 3(b)). As a result, the emitted electron in β⁻ decay is spin-polarized with only left-handed chirality. The mass and lifetime of $Z^0$ are enormous and short, respectively, making the weak force extremely short-ranged and hard to detect. However, the resulted spin-polarized electrons can be detected. The SPE is a relic particle or reporter particle from the weak interaction. After being born in and traveling outside of the nucleus, it reaches the particle detector, declaring the parity nonconservation in weak interactions [9].

The symmetry breaking is not limited to radioactive atoms and beta decays. The weak force via the Electroweak Theory renders all atomic nuclei chiral, and consequently all atoms and molecules own different degrees of chirality [15, 16]. Theoretical calculations show that there exist nonzero electron chirality densities at the position of every nucleus in a chiral molecule (e.g., amino acids) [17] or a twisted quasi-chiral molecule (e.g., $C_2H_2$, $H_2O_2$) [16-18]. The net electron chirality, which is the integrated value of the electron chirality density over the whole molecule, is also nonzero. The weak force is very slight in strength, and its action range is extremely short (less than $10^{-17}$ m). So almost all orbiting electrons in a stable atom are actually out of the range of the weak interaction. Nevertheless, the interaction between atom's orbiting electrons and quarks or nucleons via $Z^0$ in the nucleus is still possible. The electrons may wander close enough to the nucleus for the weak force to exert a more powerful influence because of the Heisenberg Uncertainty Principle. In this way, a nonzero electron chirality density occurs in every chiral molecule, as well as a parity-violating energy difference (PVED) between the two enantiomers of the molecule [16, 17].

Moreover, the asymmetric inductive effect is cumulative. The more nucleons and thus quarks an atom possesses, the larger the effect will be. Consequently, any observable effect will be more remarkable in atoms with large atomic numbers or molecules composed of atoms with large atomic numbers [15].

It is in principle possible to calculate the electron chirality bias of a chiral molecule and the PVED between its two enantiomers. Thanks to the advances in theory and computing power, the quantum mechanical calculations are today much better grounded. The PVED and the total electron chirality of L-amino acids are on the order of about $10^{-18}$ eV per molecule and $10^{-9}$, respectively. Furthermore, they are expected to be several hundred-fold larger in excited or ionized states than in ground state [17, 19]. Loosely speaking, the weak force may give rise to electron chirality and PVED at the quantum level, and then to molecular chirality. The causality between the weak force and the PVED implies that the origin of molecular chirality may be ascribed to the transfer of asymmetry from inside to outside of a nucleus, i.e., weak nuclear force → parity violation → orbiting electron chirality and PVED → molecular chirality.

However, the weak force at quark's scale is too weak and short-ranged to mediate the stereoconfiguration of a chiral molecule, which is subject to the far stronger electromagnetic force at atomic to molecular scales. Meanwhile, the PVED is extraordinarily tiny and has not yet been unambiguously detected by experiment [20]. Even though it is generally believed that the weak force renders enantiomers energetically non-equivalent, was this pervasive but feeble force responsible for the origin of biomolecules with a single handedness? This is a question of much debate [21, 22] but certainly worth much additional scrutiny.

Now let us turn the question around. Since the intrinsic weak force is not puissant enough to arbitrarily determine the stereo-asymmetry outside of its range, is there an extrinsic asymmetric factor able to modulate the molecular chirality alone on the outside, or synergistically with the slight electron chirality resulted from the inside out? That is the case of the assumed SICS mechanism (Fig. 1), in which an incident spin-polarized electron or H-atom can be deemed as the outside asymmetric factor [2, 4].

In Fig. 1, the amino acid radical has an electron-deficient carbon atom. One can usually think of an alkyl radical (e.g., methyl radical, ·CH$_3$) as planar in shape (Fig. 4). It is conveniently described by sp$^2$ hybridization, with the unpaired electron located in p-orbit and distributed symmetrically on the two sides of the plane. However, some theoretical and experimental studies have provided new insight into the geometry of these radicals. In general, the structural possibilities include planar, rapidly inverting pyramidal and rigid pyramidal structures (Fig. 4). Spectroscopic evidence shows that even the structure of the methyl radical could be either planar or a shallow pyramid with a very low energy barrier against the inversion [23]. Especially at very low temperature (15 K), the IR spectra of ·CH$_3$ indicate a maximum of about 5$^o$ on the deviation from planarity [24]. Substituted alkyl radicals become more pyramidal [25]. The tendency for distortion from planarity to pyramidal geometry results from the repulsive tension among electrons or orbits.

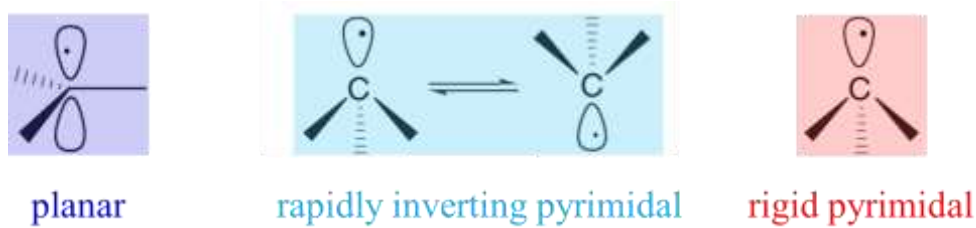

**Fig. 4** Possible stereostructures of a carbon-centered free radical.

Deductions about nonplanarity can also be drawn from the study of the stereochemistry of reactions involving radical intermediates. Generally, the order of stability of alkyl radicals is 3° > 2° > 1°. A tertiary radical is comparatively stable. If the carbon center of a radical is jointed to three different groups (e.g., an α-amino acid radical) and the unpaired electron could be visualized as a fourth bond, the radical goes from planar sp$^2$ hybridization to tetrahedral sp$^3$ hybridization. If so, the pyramidal shape seems to be a quasi-chiral or pre-chiral configuration, with the carbon atom as a stereogenic center. The product would be racemic if the radical is a planar or rapidly inverting one, whereas a rigid pyramidal structure would lead to product of retained handedness. Some model reactions have been subjected to careful examination of these concerns, but racemic product is formed in each case, indicating the quasi-chiral precursor radicals do not retain their tetrahedral geometry [26-29]. But on the other hand, the inversion of the pyramidal structure is also subject to molecular constitution and conformation. Alpha-amino acids are zwitterionic and able to form five-membered ring structures in nonplanar envelope-like conformation, via intramolecular hydrogen bonding (Fig. 5). This structure is expected to have a higher barrier against the radical inversion, making the stereoselective synthesis more possible. For cyclohexyl radical as an example, an $E_a$ of 5.6 kcal/mol has been measured for conformation inversion [30].

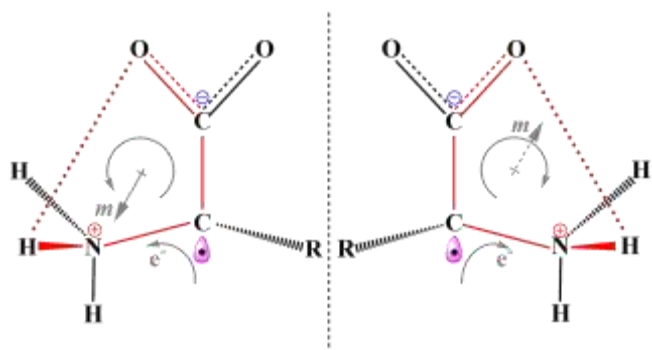

**Fig. 5** Zwitterionic α-amino acid radical enantiomers stabilized by intramolecular hydrogen-bonding.

The amino acid radical is quasi-chiral at the carbon center. Hence it has a nonzero local electron chirality density around the carbon nucleus. Moreover, the amino acid radical has a protonated amino group, a deprotonated carboxylic group and an unpaired electron. From the viewpoint of whether chemistry or physics, it exists in ionized state. Therefore, the electron chirality density is supposed to be much larger since the radical is in open shell ionized state [17, 19].

A radical is usually unstable and highly reactive due to its incompletely satisfied valency. It tends to abstract electrons from outside or couple with another radical. Now let us switch focus to 'another radical', the SPE or the spin-polarized H-atom (SPH) in the SICS reaction (Fig. 1). When the SPH approaches the electron-deficient carbon center of the amino acid radical, a new C-H σ bond is formed by the interaction of C-$sp^3$ with H-1s orbits. The H-1s electron is completely spin-polarized, with the spins of the electron and nucleus in opposite directions (antiparallel) [2, 4]. This structure of the SPH is lower in energy and more stable than its counterpart with the spins of the electron and nucleus in the same direction. The partner C-$sp^3$ electron is also polarized, though very slightly. Or rather, it has a very low asymmetric electron chirality density at the α-C site. During the coupling process of the two radicals, there would be a preferential attack direction for the SPH. According to the Pauli Exclusion Principle, It tends to proceed to the region where the spin of the unpaired C-$sp^3$ electron is antiparallel to the spin of its own 1s electron. Since the spatial distribution of the spin state of the $sp^3$ electron is asymmetric, the radical coupling reaction in Fig.1 will bring forth amino acids with determined chirality.

As mentioned earlier, the electron chirality density or the PVED resulted from the weak force is too small to regulate the chirality of the product (here named internal determinism). And currently there is no clear evidence for a causal connection between the PVED and biomolecular homochirality. The spin induced chiral selectivity may be mediated by the spin state of the incident particle (external determinism), or co-determined by the spin states of both the incident particle and the molecular radical (reciprocal determinism). If reciprocally, it should be noted the SICS conjecture has one fundamental drawback: the electron chirality density and PVED in the molecular radical are extremely limited. How is it possible to make the tiny differences into something meaningful where the daunting homochirality is possible? It could be imagined that there exists a *spin-inductive effect* when the completely spin-polarized electron or H-atom gets closer to the unpaired C-$sp^3$ electron. The spin chirality density of the C-$sp^3$ electron could turn significantly larger through spin-spin interaction between the SPH and itself. By this iterative amplification mechanism, the small asymmetry in the latter gets amplified and then the chiral synthesis becomes compulsive.

Additionally, in the case of amino acid, the approach of the electron or H-atom to the α-C will generate an oscillating electronic current in the five-membered ring which is closed by hydrogen bonding (Fig. 5). The ring current gives rise to a large magnetic dipole transition moment (*m*) which has been experimentally observed in enantiomers underlying their opposite optical rotations [31]. The inductive magnetic field will conversely exert an influence on the $sp^3$ electron, leading to its spin antiparallel to that of the incident electron. This is called *magnet-inductive effect*.

Taken into consideration all the electromagnetic, spin-spin, spin-orbit coupling and weak force interactions, the SICS mechanism is able to be described in more details by quantum electrodynamics (QED) and Electroweak Theory. In essence, chemical reaction is the redistribution of electrons or the reconstruction of electronic wavefunctions. The chemical reaction can be visualized using electron density, electron chirality density and electronic energy density. The electronic energy density is a unique function of the electron density and electron chirality density. From the viewpoint of energy, the energy density is the driving force of chemical reaction. The three densities can identify the intrinsic shape of the reactants, transition states, and products, along the course of the chemical reaction coordinate [32-34]. That is to say, the formation of a chemical bond can be iconically visualized in real 3D space [35].

The total energetic density is the sum of the kinetic energy density, the external potential energy density (including the featureless Hartree part), and the interelectron potential energy (exchange-correlation effects). In classic mechanics and electromagnetics, we know that force is the negative gradient of potential energy which acts on an object or a charged particle to make it move. Similarly, in a quantum system there exists a quantum force, the relativistic stress/tension force, as well as the Lorentz force [32-34]. Interestingly, this force can help realize the regioselectivity of chemical reactions [35, 36]. In addition, the stress force is spin-dependent. The spin-related part of the force is called spin torque [37-39]. The spin torque is completely canceled out by zeta force in a steady state atom, but possibly leaving a nonzero net torque in a nonsteady state atom. The net spin torque, just like the chirality density, is expected to be a residual effect of the weak force (negligible) and the spin–orbit interaction. But the magnitude of the net spin torque is expected to be not as small as the chirality density, since it results from electromagnetic interaction which is far stronger than the weak interaction. It is even more so for the molecular radical in the SICS reaction (Fig.1), because the radical occurs in open shell ionized state. Meanwhile, the spin torque of the incident SPE or SPH is presumed to be very strong. During the radical coupling reaction, it may act as a guiding force for the synthesis of chiral molecules through spin torque destruction and reconstruction. That is to say, the asymmetric stress, spin torque and their asymmetric spatial distribution are the clinchers determining the emergence of molecular chirality.

Hence, by using the SPE or the SPH as a chiral reagent, the SICS pathway might open a door to exploiting (electrocatalytic) radical reactions not only to control the stereochemistry of the product, but also to increase the scope of suitable reaction partners for its wider applications in chiral synthesis.

## Experimental evidence

The weak force, which is involved in nuclear decay, is the only force of nature already known to have a handedness preference: electrons created in β⁻ decay are always left-handed. Hence, immediately after the discovery of the parity violation in weak interactions in 1956-1957 [8, 9], it has been speculated the parity-violating weak neutral current mediated by $Z^0$ boson at the molecular level could be a potential answer to the origin of bio-homochirality [40, 41]. This preconception has later been assessed by a variety of experimental designs, using SPEs produced in β⁻ decay or by accelerator to induce asymmetric decomposition of racemates [42-47]. However, all those attempts to confirm the effect have been frustrated because they failed to detect any undoubted enantiomeric excesses and even there were outright contradictions among those results. Instead of enriching one enantiomer over the other, the reaction initiated by beta particles on the very contrary tends to racemize the chiral molecule [48]. In 2000, Bonner summarized the situation up to the time by concluding that the symmetry-breaking in biomolecules was not likely to be the result of parity violations in the weak interaction [22].

In the last 15 years, with the development of spintronics, some new experimental methods and skills have been applied to the investigation on the causal relationship between electron spin and molecular chirality. Naaman *et al.* found that achiral photoelectrons ejected from gold substrate become polarized after transfer from one side to the other side through a self-assembled monolayer of double-stranded DNA on the substrate [49]. The DNA acts as a spin filter to realize spin-selective transport of the photoelectrons and then create SPEs. This phenomenon, called *chiral-induced spin selectivity* (CISS), has also been observed in other chiral molecules such as α-helical oligopeptides and proteins [50]. Circularly polarized light can also produce SPEs. By combining those techniques, Rosenberg *et al.* [51, 52] fired the low-energy SPEs from magnetic alloy or DNA arrays at chiral molecules, resulting in the falling apart of the molecules. By measuring how often the reaction occurred for each handedness of the electron and molecule, they found that left- or right-handed electrons preferentially destroy one version of a molecule over its mirror image, i.e., *spin-induced asymmetric decomposition*. The enantiomeric excesses (ee) were about 9-16% [51, 52]. However, it appeared that the reaction initiated by SPEs that enriches one enantiomer over the other also synchronously ruined the molecules as well. It was estimated that when the ee was 25%, almost 99% of the initial substrate would be destroyed [51]. The subject how SPEs could transfer their asymmetry to molecules needs an alternative constructive but not destructive perspective.

In 2017, by using a chiral molecule decorated anode, Naaman and his colleagues demonstrated that SPEs can enforce thermodynamic constraints on the electrocatalytic water splitting reaction, yielding one product species preferentially over the other [53]. The reaction mechanism is summarized as below,

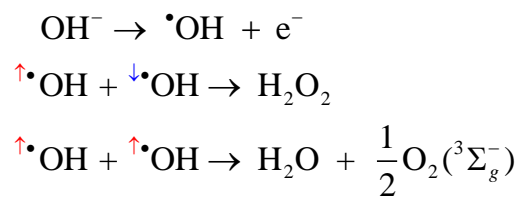

$$OH^- \rightarrow {}^\bullet OH + e^-$$
$$\uparrow{}^\bullet OH + \downarrow{}^\bullet OH \rightarrow H_2O_2$$
$$\uparrow{}^\bullet OH + \uparrow{}^\bullet OH \rightarrow H_2O + \frac{1}{2}O_2({}^3\Sigma_g^-)$$

When a bare or achiral molecule decorated anode is used, the electrons transfer from hydroxide ions to the anode is non spin specific and the spins of the remnant unpaired electrons on the •OH radicals are racemic (aligning antiparallel to each other), hence the two •OH combine to produce $H_2O_2$. When a chiral anode is used, the spins of the unpaired electrons of •OH radicals are polarized (aligning parallel), hence the radical coupling reaction is forbidden, facilitating the production of oxygen in its triplet ground state. This can be called *spin-induced product selectivity*.

A third pioneering research by Naaman's group is about *spin-induced asymmetric adsorption* of chiral molecules on magnetic materials. The electron has two intrinsic properties, charge and spin. As well as electron density, there is also electron chirality density, i.e., asymmetric spatial distribution of electron spin, in a chiral molecule. When the chiral molecule gets closer to another one or to a magnetic substrate, a chiral recognition event will take place between them. Both the electron density and electron chirality density will be redistributed in each one. Charge polarization is accompanied by spin polarization for chiral molecules. The spin polarization, arising from spin-spin interactions and spin-inductive effect, may constrain the symmetry of the wave functions and play an important role in the molecular recognition process [54]. This spin-dependent enantioselectivity has also been observed in the interaction of chiral molecules with achiral magnetic inorganic substrate [55]. One enantiomer of cysteine or oligopeptides adsorbs onto the substrate of a certain magnetized state preferentially over the other, resulting in the separation and enrichment of enantiomers.

More fascinatingly, in 2020, Naaman *et al.* further demonstrated that SPEs can drive asymmetric synthesis of chiral molecules [56], that is, *the spin-induced chiral selectivity* suggested as early as 2013 [2-4]. Circular dichroism spectra showed that both the reduction product of camphorsulfonic acid (i.e., mercaptoborneol) and the oxidation product of 1-pyrenecarboxylic acid (i.e., pyrene polymers) formed on the surface of magnetized nickel electrodes were chiral. The enantiomeric excess of the mercaptoborneol was about 11.5%. The enantioselectivity did not result from enantiospecific interactions (e.g., spin-induced asymmetric adsorption) of the molecule with the ferromagnetic electrode but from the SPEs injected by the magnetic metal. This means that SPE can act as a "chiral reagent" to induce stereoselective synthesis of chiral molecules.

In fact, three years earlier, the author's group has also observed the SICS phenomenon in the electrocatalytic reductive amination of α-oxo acids to α-amino acids, by using $Fe_{0.2}Ni_{0.8}$ thin films as the cathode [57]. We obtained a maximal ee of 10.4% of the amino acid products. But the repeatability was not good. The ee values were only 2-5% in most cases. The reason for this was considered to be that the surface morphologies of the films have an obvious influence on the ee value. When the film was grown to a thickness larger than 1 μm, it exhibited a needle-like and porous nanostructure at the top (Fig. 6). The surfaces looked black but not bright silver to the naked eye, due to their low reflectance over a broad spectral range and pronounced light trapping effect. This is just like the case of black silicon [58]. The sharper the needles, the blacker the color, and the higher the ee. The higher ee may be ascribed to the higher transfer efficiency of the spin-polarized electric current through the needles, via an unknown mechanism.

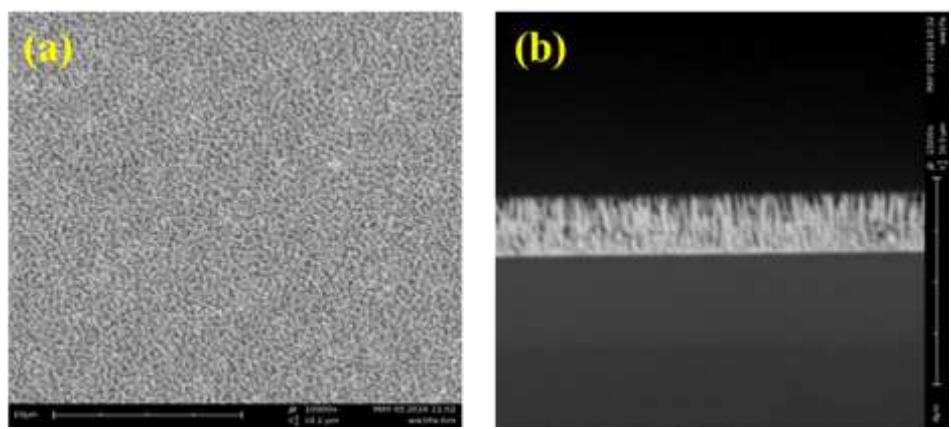

**Fig. 6** SEM images of the surface of a $Fe_{0.2}Ni_{0.8}$ thin film. (a) top view, (b) side view.

The above-mentioned experimental evidence in recent years indicates that, 1) chiral molecules can create SPEs; 2) SPEs can induce the asymmetric adsorption, decomposition, recognition, and synthesis of chiral molecules; 3) the reciprocal interactions between SPEs and chiral molecules via the CISS and SICS mechanisms can realize the induction, transfer and generation of the two asymmetries. Phenomenologically speaking, there is an underlying *cross-scale* relationship between electron spin and molecular chirality (Fig. 2).

That SPEs could transmit their asymmetry to organic molecules makes the SICS mechanism very attractive, because it might help explain the origin of the bio-homochirality. However, to the present there is still a lack of appreciation of what is actually going on in the SICS process. More theoretical and experimental evidence is needed to uncover the details in the process. According to the force-motion philosophy, there should be an asymmetric spatial distribution of an unknown force which results in asymmetric cross sections for the collision of the two radicals, and then the stereoselective formation of one enantiomer. The QED stress force [32-34] or spin torque [37-39] might be such a force candidate. It arises from electromagnetic interactions which is about eleven orders of magnitude stronger than the weak force. Recently, Chen *et al.* reported an experimental study on the gas-phase hydrogen atom transfer reaction, i.e., F + HD → HF + D, by using high-resolution imaging techniques [59]. The authors observed a peculiar horseshoe-shaped imaging pattern of the D-atom product on the rotation state-resolved differential cross sections. The scattering profiles reflect not only the angular distribution of the speed of D-atoms (H-atom donor product), but also the resonance states of HF (H-atom acceptor product). To explain the unusual reaction dynamic pattern, they performed quantum dynamics calculations by using a full open-shell spin-orbit coupling model, which included *all* angular momentum couplings between electronic spin, electron orbital angular momentum, and nuclear rotational angular momentum. The theoretical analyses showed that the complicated couplings lead to the splitting of the single partial wave of HF into the fine structure of four partial wave resonances. The quantum interference between partial wave resonances with positive and negative parities resulted in the horseshoe-shaped pattern. Notably, the theoretical results were in near perfect agreement with the spatial scattering distribution of the products manifested in the imaging pattern. So this contribution excellently demonstrates that spin and spin-orbit interaction can effectively influence reaction dynamics. In light of this thought-provoking work, I would entertain a notion that if the unpaired 2p electron of the incident F-atom is spin-polarized, is it possible that the horseshoe-shaped pattern turns into a symmetry-broken structure (a half horseshoe), i.e., an asymmetric spatial distribution of the products?

## Concluding remarks

The origin of bio-homochirality is a subject of much debate, still remaining a giant puzzle. The origin of life on earth may be a break of symmetry to be compared with the breaks of symmetry in the evolution of the universe. The evolution of the universe in general and life in particular can be quantitatively considered as a cascading process involved with a series of symmetry breaking. During the process, the emergence, transfer, and transformation of different asymmetries took place one by one. The biological chirality might arise from the cross-scale transfer of asymmetries with different forms at various scales, from elementary particles to atoms, micromolecules, macromolecules, ending up in biological entities (Fig. 2). Based on the perspective of asymmetry transfer, the molecular chirality may stem from the well-known electron spin at subatomic level, i.e., the spin-induced chiral selectivity.

In this paper, the SICS mechanism and its outreach are discussed in details. A preliminary conclusion is that the QED stress force or spin torque might be the driving force for the transfer of asymmetry and the formation of molecular chirality. Some recent interesting experimental evidence seems to support the SICS conjecture. If spin-polarized electrons caused life to become chirally selective, what would have produced those electrons in the first place? A magnetic half-metal material such as greigite ($Fe_3S_4$), a mineral present in a possible primordial site where life could have arisen, maybe have acted as a spin filter to produce SPEs [4]. The resulted SPEs then induced the asymmetric synthesis of chiral molecules via the SICS mechanism.

X. Quantum interference between spin-orbit split partial waves in the F + HD → HF + D reaction. Science, 2021, 371: 936-940.